# Self-Calibrated Atom-Interferometer Gyroscope by Modulating Atomic Velocities

Hong-Hui Chen,[1, 2] Zhan-Wei Yao,[1, 3, *] Ze-Xi Lu,[1, 2] Si-Bin Lu,[1] Min Jiang,[1] Shao-Kang Li,[1] Xiao-Li Chen,[1, 2] Chuan Sun,[1, 2] Yin-Fei Mao,[1, 2] Yang Li,[1, 2] Run-Bing Li,[1, 3, 4, †] Jin Wang,[1, 3, 4] and Ming-Sheng Zhan[1, 3, 4]

[1]*State Key Laboratory of Magnetic Resonance and Atomic and Molecular Physics,*
*Innovation Academy for Precision Measurement Science and Technology,*
*Chinese Academy of Sciences, Wuhan 430071, China*
[2]*School of Physics, University of Chinese Academy of Sciences, Beijing 100049, China*
[3]*Hefei National Laboratory, Hefei 230088, China*
[4]*Wuhan Institute of Quantum Technology, Wuhan 430206, China*
(Dated: February 14, 2023)

Atom-interferometer gyroscopes have attracted much attention for their potential superior long-term stability and extremely low drift. For such high precision instrument, a self-calibration to achieve an absolute rotation measurement is highly demanded. Here we propose and demonstrate a self-calibration of the atomic gyroscope. The calibration is realized by using the detuning of laser frequency to control the atomic velocity thus to modulate the scale factor of the gyroscope. The modulation determines the order and the initial phase of the interference stripe, thus eliminates the ambiguity caused by the periodicity of the interferometric signal. The calibration method is verified by measuring the Earth's rotation. Long-term stable and self-calibrated atom-interferometer gyros can find important applications in the fields of fundamental physics and long-time navigation.

Rotation measurement with inertial sensors plays an important role in the fields of fundamental physics [1], geophysics [2] and inertial navigation [3]. For example, to measure the Lense-Thirring frame dragging or other general relativistic precessions, absolute rotation sensors with extreme sensitivities are urgently demanded [4–8]. Atom-interferometer gyroscopes are a class of quantum sensors that have a huge potential for realizing a high-sensitivity rotation measurement [9–14]. They were rapidly developed in the past two decades, and the sensitivity was well improved by enlarging the scale factor and by suppressing phase noises [15–27]. However, like any interferometric signal due to the periodicity and the bias phase, it is difficult to determine the absolute inertial phase shift. The periodicity causes an ambiguity of the phase determination because the rotation induced phase shift may exceed $2\pi$ radians as the scale factor increased or in real-world applications [28]. To determine an absolute inertial phase shift, calibrated schemes with different sensors, such as classical and matter-wave hybridized sensors [29–31], and multi-species composite-fringe atom interferometers [32–34], were proposed.

It is the potential superior long-term stability and extremely low drift that makes the atom-interferometer gyroscopes be candidates for next generation gyros. To realize an absolute rotation measurement, both of the scale factor and the rotation induced phase shift should be calibrated as accurately as possible. Recently, the Earth rotation was measured and also used to calibrate the atom gyroscope with an auxiliary turntable [14]. In this con-dition, the Earth's rotation rate projected on the orientation of the gyroscope is modulated as the turntable rotated. This procedure induces extra systematic errors due to the geomagnetic field or the wobble of the turntable, which will limit the improvement of the precision. On the other hand, for the higher precision atom gyroscope, it needs to be mounted on the massive rock like the laser gyroscope to reduce the environment noise [4], so no other instrument can calibrate it. Thus, how to make a self-calibration of atom gyroscope to achieve absolute rotation measurement is an unfulfilled task.

In this paper, we propose and demonstrate a self-calibration of a large area atom-interferometer gyroscope by modulating atomic velocities. The calibration is realized by using the frequency detuning of cooling lasers to control the atomic speed thus to modulate the scale factor of the gyroscope. The modulation determines the order and initial phase of the interference stripe, thus eliminates the ambiguity caused by the periodicity of the interferometric signal. This idea is novel in Sagnac effect based gyroscopes, compared with the similar optical (laser and fiber) gyroscopes where the light speed is unchangeable. This method has more advantages, compared with the interrogation time modulated, because the position-dependent phase shift is much reduced. In the following, we explain how to develop and verify this methodology in a dual-atom-interferometer gyroscope.

The scale factor ($K$) is defined as the ratio between the rotation induced phase shift ($\phi_\Omega$) and the rotation rate ($\Omega$), namely $K = \phi_\Omega / \Omega$. In the dual-atom-interferometer gyroscope, when the dual interferometric loops are constructed by three pairs of Raman pulses as shown in Fig.1 (a), the differential phase shift caused by the rotation be-tween two atomic interferometers is given by

$$\phi_\Omega = 4\vec{k}_{\rm eff} \bullet (\vec{\Omega} \times \vec{v})T^2, \qquad (1)$$

where, $\vec{\Omega}$ is the rotation in the inertial frame, $\vec{k}_{\rm eff}$ is the effective wave vector of Raman lasers, $\vec{v}$ is the atomic velocity, and $T$ is the interrogation time between two adjacent Raman pulses. When the duration effect of Raman pulses is considered [35] and the rotation is projected on interferometric loops, the scale factor is written as

$$K = 4k_{\rm eff}vT^2[1 + (\frac{4}{\pi} - 2)\frac{\tau}{T}], \qquad (2)$$

where, $\tau$ is the $\pi/2$-pulse duration of Raman lasers, and $v$ is the atomic velocity along the horizontal direction. As shown in Fig.$1$ (b), by definition the rotation induced phase shift is proportional to the scale factor(red solid line), $\phi_\Omega = K\Omega$, but the measured differential phase shift ($\phi_D$) is a periodic function (red dashed line), as given by

$$\phi_D = \phi_\Omega - 2n\pi + \phi_0, \qquad (3)$$

where, $n$ is an integral number and $\phi_0$ is an initial phase. For a fixed rotation rate $\Omega$, by calibrating and modulating the scale factor $K$, $\phi_D$ are measured and $n$ can be determined by the intercept of the linear fitting line. Thus, $\phi_\Omega$ is obtained from Eq.($3$), and the absolute rotation rate is given by $\Omega = \phi_\Omega/K$. Theoretically, the scale factor can be modulated by controlling the atomic velocity or the interrogation time. However, from Fig.$1$ (a), a position-dependent phase shift will be induced when the interrogation time is modulated, because the atoms are moved not parallel to the Raman beams in the atom-interferometer gyroscope. To avoid this effect, the scale factor is modulated by changing the atomic velocity for a fixed interrogation length between two Raman pulses. Furthermore, it is clear that the scale factor $K$ can be calibrated by precisely determining the atomic horizontal velocity according to Eq.($2$). We call this method as velocity-modulated linear fitting method (VMLF).

The experimental apparatus used to demonstrate VMLF, as shown in Fig.$1$ (a), was mainly described in the previous work [21]. Briefly, the dual-atom-interferometer gyroscope consists of two symmetric atom interferometers. The $^{87}$Rb atoms are loaded in two magneto-optical traps (MOTs) and then launched along two symmetric parabolic curves with opposite horizontal velocities by the optical moving molasses technique. The atom number is more than $1 \times 10^7$ and its dimension is 3 mm. The atomic temperature is less than 14 $\mu$K and its corresponding velocity distribution is 3.7 cm/s. Two Mach-Zehnder interferometric loops are constructed by three pairs of Raman pulses, and interference fringes are observed by laser-induced fluorescence signals. The cycle time is 0.25 s for each interference measurement. The differential phase shift is extracted from two atom interferometers. Furthermore, the scale factor, which is proportional to the interference-loop area embraced by three pairs of Raman pulses, is modulated by precisely controlling the horizontal atomic velocities as proposed above. The separation between two adjacent Raman beams is fixed at about 20 cm. The frequency detunings of three pairs of cooling lasers are independently controlled. The atomic horizontal velocity is controlled by the frequency detunings of two pairs of cooling lasers. When horizontal atomic velocities are changed, the atomic trajectories are slightly changed, as shown by the blue solid line, red dashed line, and green dot-dashed line in Fig.$1$ (a). To make two atomic interferometric loops overlap, the vertical atomic velocities are modified by adjusting the frequency detuning of the third pairs of cooling lasers.

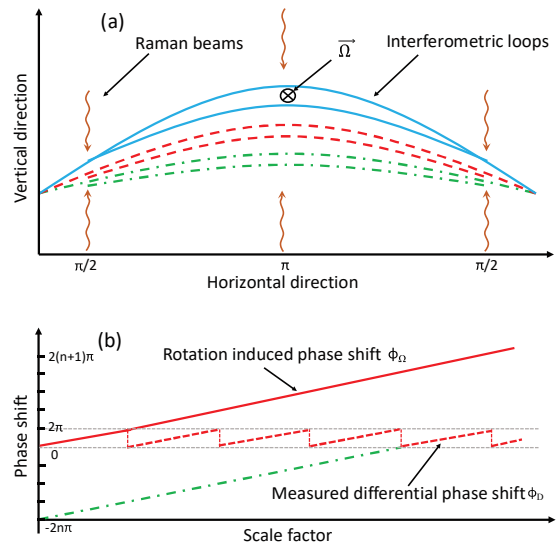

FIG. 1: (color online) Schematic diagrams of a dual-atom-interferometer gyroscope (a) and the velocity-modulated linear fitting method (b). Three pairs of Raman beams are applied to construct dual Mach-Zehnder atom interferometers. The atomic trajectories are changed as shown in different parabolic curves when the scale factor is modulated by controlling the atomic velocity. The measured differential phase shift ($\phi_D$) is extracted by the sine fitting method, and the rotation induced differential phase shift ($\phi_\Omega$) is determined by the linear fitting method with the intercept considered.

Now we show how to calibrate the scale factor. In Eq.($2$), as the wave vector (frequency) and time can be measured and controlled precisely enough, the atomic velocity becomes major challenge to accurately determine the scale factor. We develop a two-slit Ramsey interferometer method for measuring an absolute atomic velocity, in which Raman beams with two slits are applied for constructing a Doppler insensitive Ramsey interferometer. When the distance $L$ between two slits is fixed, the atomic velocity can be precisely determined by measuring the phase difference between two Ramsey fringes with different two-photon detunings of the Raman lasers, i.e.

$$v = \frac{L}{T_R} = \frac{2\pi L(f_1 - f_2)}{\phi_1 - \phi_2}, \qquad (4)$$

where, $T_R$ is the interrogation time between two slits, $f_i(i = 1, 2)$ are the two-photon detunings of the Raman lasers and $\phi_i$ are the phases of the Ramsey fringes for both measurements. In the experiment, Ramsey interferometers are built by the first and second pairs of co-propagating Raman beams whose widths are narrowed to 900 $\mu$m. Narrowed Raman beams can broaden the envelope of Ramsey fringe and improve the precision of the interrogation time $T_R$. The interrogation time is measured from the phase difference ($\phi_1 - \phi_2$) and frequency difference ($f_1 - f_2$ ), where the whole envelope of Ramsey fringe is scanned over a frequency range of 5.8 kHz to determine the period number and then the phase shift is extracted by the mid-fringe method when the frequency interval is set to about 2 kHz. The atomic velocities are measured with the two-slit Ramsey interferometer method for various frequency detunings of cooling lasers, as shown in Fig.2 (a). The experimental data (black dots) show that the atomic velocity can cover a range from 3 to 6 m/s in our experimental setup. The minimum velocity is limited by the height of the inner chamber. The residual errors between the experimental data and the fitting results are shown in Fig.2 (b). To determine the scale factor, the fluctuation and systematic error of the atomic velocity are evaluated when the frequency detuning is fixed at 4.0400 MHz. In this case, the frequency interval is set to 2007.200 Hz confirmed by a frequency spectrum analyzer with a resolution of 1 mHz. The phase difference is extracted from the period number and the phase shift as mentioned above, and it is 581.115 rad with an uncertainty of 40 mrad. With the help of a motorized translation stage, $L$ is measured to 203.693 mm with an accuracy of 20 $\mu$m. Put all of the errors together in Table I, the relative uncertainty of the horizontal velocity is 122 ppm. and its magnitude is evaluated to $4.4206 \pm 0.0005$ m/s. The stability of the atomic velocity is less than 4 ppm at the integration time of thousands of seconds.

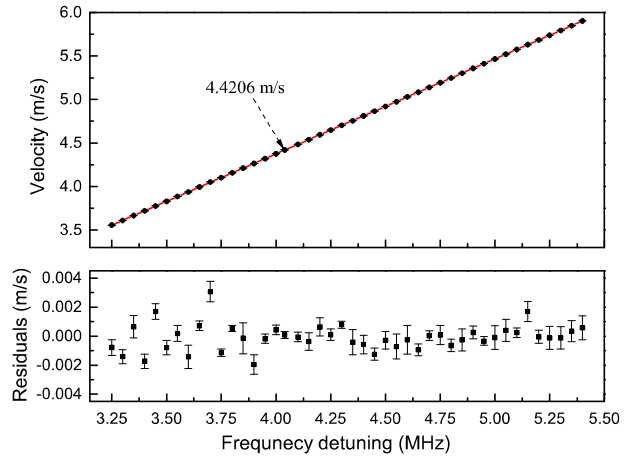

FIG. 2: (color online) Dependence of the atomic velocity on the frequency detuning of cooling lasers (a) and residual errors between the experimental data and the fitted results (b). Atomic velocities are measured with the two-slit Ramsey interferometer method. The black dots and squares are experimental data, and the red line is the fitting curve.

TABLE I: **Error sources of horizontal atomic velocity measurements**. The table shows the uncertainties of the parameters to measure the horizontal atomic velocity. The relative uncertainties are determined by the measurement values and their errors. The horizontal atomic velocity is evaluated when the contribution of all parameters is considered.

| Parameter | Value (Unit) | Relative uncertainty (ppm) |
|---|---|---|
| Frequency | 2007.200 (1) Hz | 0.5 |
| Phase | 581.115 (40) rad | 70 |
| Length | 203.693 (20) mm | 100 |
| Atomic velocity | 4.4206 (5) m/s | 122 |

After the atomic horizontal velocity was calibrated, the slits are removed, and a dual-atom-interferometer gyroscope is built with two symmetric Doppler sensitive atom interferometers by using three pairs of counter-propagating Raman beams as shown in Fig.1 (a). To enhance the atom coherence, the atomic velocity distribu-tions are reduced to 12.5 mm/s by using velocity-selected Raman transitions. When the atomic velocity is 4.4206 m/s and the interrogation time is 45.810 ms, two interference fringes are observed with contrasts of 25.6% and 20.5%, as shown in Fig.3 (a) and (b). Black squares and red dots are experimental data, and solid lines are fitted results by the sine function. Compared with our previous work [21], fringe contrasts are improved by lowering the atomic temperature and by optimizing the Raman-pulse width. For different atomic velocities, two interference fringes can be obtained, and the phase shift for each of two interferometers is extracted by the sine fitting method. The phase uncertainty is less than 30 mrad for each fringe. The measured differential phase shift between two interferometers, corresponding to the rotation term, are extracted for different atomic velocities, as shown in Fig.4 (a), which is a periodic function. However, atomic trajectories are slightly changed as shown in Fig.1 (a) when their velocities are modulated. The misalignment of Raman beams produces a position-dependent phase shift, and an error is introduced when differential phase shifts are extracted in dual atom interferometers. This error is originated from the coupling between the misalignment of the Raman lasers and the position displacement of atoms, and was reduced by optimizing the timing sequence and by aligning the Raman beams [36].

To calibrate the rotation induced phase shift $\phi_\Omega$, we show how to scan the factor to determine $n$ at certain fixed rotation rate $\Omega$. In Fig.4, the orientation of the gyroscope is fixed at an arbitrary azimuth angle ($\alpha$) with respect to the north direction. The atomic velocity is precisely controlled and scanned by the frequency detuning of cooling lasers as measured in Fig.2 (a). The scale factors are correspondingly obtained by Eq.(2). When the scale factor is scanned from 2.0 to 3.6 rad/(°/h), the dependence of the measured differential phase shift ($\phi_D$), which is the phase difference between two interference fringes, is shown in Fig.4 (a). It can be seen that the measured differential phase shifts induced by Earth's rotation cover three periods, i.e., black squares, red dots and blue triangles (experimental data), and they are linearly fitted (solid lines), respectively. The fitted results show that the slope is 10.310±0.052 °/h and the intercept is −19.002±0.163 rad when the scale factor is modulated from 2.0 to 2.4 rad/(°/h), the slope is 10.291 ± 0.049 °/h and the intercept is −24.901±0.135 rad when modulated from 2.5 to 3.0 rad/(°/h), and the slope is 10.295±0.045 °/h and the intercept is −31.364±0.143 rad when modulated from 3.1 to 3.6 rad/(°/h). Theoretically, the slope is the Earth's rotation rate projected on interferometric loops and the intercept corresponds to $2n\pi + \phi_0$ in Eq.(3). Actually, in the dual-atom-interferometer gyroscope, $\phi_0$ is commonly canceled. From the intercept of the linear fitting lines, the values of $n = 3, 4, 5$ are derived for the three periods. Therefore, once $n$ is decided, the one-to-one correspondence between $\Omega$ and $\phi_\Omega$ is established. However, the correspondence is still constrained by the residual statistical error between experimental data and the fitted results as shown in Fig.4 (b), as well as systematic errors due to the residual bias phase ($\phi_0$).

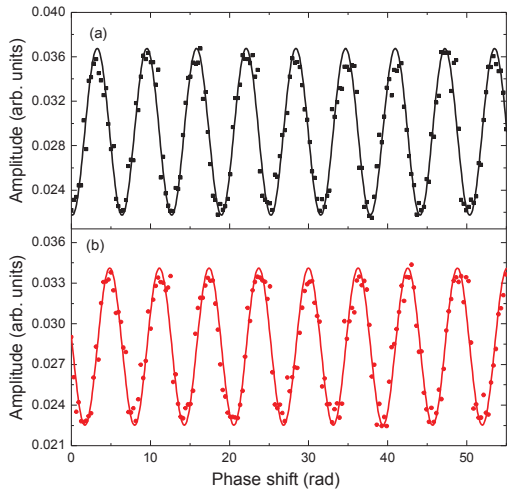

FIG. 3: (color online) Mach-Zehnder fringes of the dual atom-interferometer gyroscope. The contrast is 25.6% for the first atom interferometer and 20.5% for the second one, respectively. Black squares and red points are experimental data. Black and red solid lines are the sine fitting curves.

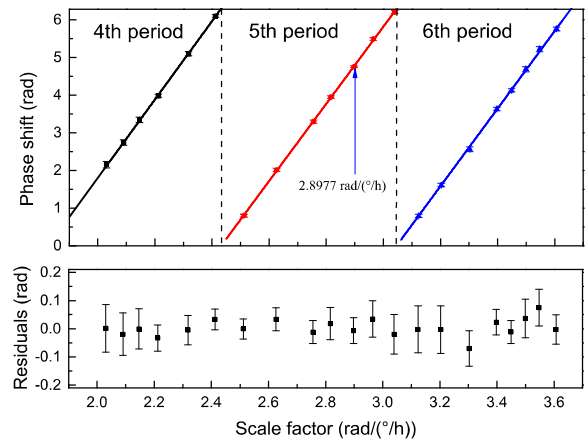

FIG. 4: (color online) Dependence of measured differential phase shifts on the scale factor (a) and residual errors between experimental data and the fitted results (b). Differential phase shifts are extracted with dual interference fringes, and scale factors are obtained by horizontal atomic velocities.

To quantitatively evaluate the magnitude of the errors, we fix the scale factor at 2.8977 rad/(°/h) in Fig.4 (arrow pointing). Firstly, to cancel the systematic error ($\phi_0$), the wave vector reversal method is applied and the overlapping of two interferometric loops is optimized by adjusting atomic velocities along the vertical direction

[18, 19]. Then, error sources are analyzed. The statistical error is 3 mrad at the integration time of thousands of seconds, corresponds to 100 ppm. Due to the residual misalignment of Raman beams, the phase uncertainty caused by the wave front of Raman lasers is 23 ppm. The gravity gradient induced error is 17 ppm due to the position mismatch along the vertical direction. Thanks to the symmetry of the dual-atom-interferometer gyroscope, the differential phase shift is insensitive to a constant Zeeman shift, but the magnetic field gradient causes a phase uncertainty. The magnetic field gradient is 14 mG/m in the horizontal direction, and causes an uncertainty of 21 ppm. The uncertainty caused by ac Stark shift is 7 ppm when considering the power fluctuation of the Raman lasers. Put all of the errors together the relative uncertainty of the absolute phase shift is 106 ppm, as given in Table II. Because the relative uncertainty of the scale factor is 122 ppm when the horizontal atomic velocity is fixed at 4.4206 m/s according to Table I [37], the relative uncertainty for an absolute rotation measurement is 162 ppm. Finally, with the self-calibrated method, the Earth's rotation projected on the orientation of atomic interferometric loops is measured to 10.3148±0.0017 °/h. With the Earth's rotation rate of 15.0410 °/h from the International Earth Rotation Service and the local latitude of 30.5417 °/h measured by the GPS signal in our lab, the orientation of the gyroscope with respect to the truth north direction, $\alpha = 37.2266 \pm 0.0062$ °, is gotten.

The absolute rotation measurement is verified by monitoring the Earth's rotation with a turntable as in the recent work [14]. The gyroscope is put on the top of a turntable, the sense of the table rotation is parallel to the Raman lasers. The Earth's rotation is projected on dual atomic interferometric loops by changing the ori-entation angle of the gyroscope. When the gyroscope is located in Wuhan and the turntable angle is changed from $-30$ to $210$ °, the dependence of Earth's rotation caused phase shift on the angle of the turntable exhibits a sinusoidal function, as shown in Fig.5 (a). Black solid dots are the experimental data, and they are sinusoidally fitted (black solid line). The orientation of the gyroscope with respect to the north direction is obtained by recording the angle of the turntable. When the turntable is located at $180.85 \pm 0.09$ ° (black dashed line), the phase shift caused by the Earth's rotation is maximum, which corresponds to the truth north direction. Because the angle of the turntable was fixed at $143.63$ ° (black dot-dashed line) when the absolute rotation rate was measured in Fig.4, $\alpha = 37.22 \pm 0.09$ ° is also obtained by the turntable-modulated method, which is consistent with $\alpha = 37.2266 \pm 0.0062$ ° gotten by the self-calibration method. From Fig.5 (a), the projection of the Earth's rotation in the truth-north direction is obtained to $12.954 \pm 0.014$ °/h. With the turntable-modulated method, the relative uncertainty is 1081 ppm for the absolute rotation measurement. Meanwhile, our gyroscope is transportable [38], and the absolute rotation was also measured with two methods in Beijing, e.g., the experimental data with turntable-modulated method are shown in Fig.5 (b). Therefore, our results prove that the self-calibrated method is valid and usable because it has higher accuracy for measuring the absolute rotation and searching the azimuth angle than the turntable-modulated method for the same experimental setup.

TABLE II: **Error sources of differential phase measurements.** The table lists the error sources for extracting the differential phase shift of the dual atom interferometers.

| Error source | Relative uncertainty (ppm) |
|---|---|
| ac Stark shift | 7 |
| Zeeman shift | 21 |
| Gravity gradient | 17 |
| Wave front | 23 |
| Statistical error | 100 |
| Total | 106 |

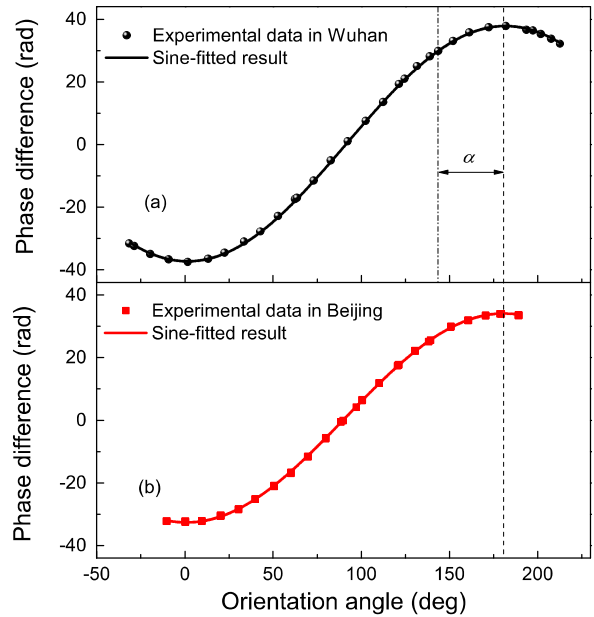

FIG. 5: (color online) Measurements of the Earths rotation rate as a function of the orientation of the dual-atom-interferometer gyroscope in (a) Wuhan and (b) Beijing. The differential phase shift depends on the sensor orientation angle in laboratory coordinates, and it is a sinusoidal function.

Furthermore, the true-north direction was measured with the turntable-modulated method in Wuhan and Beijing, and their uncertainties are $0.09$ ° and $0.12$ ° from Fig.5 (a) and (b), respectively. This implies that the north seeking system was built by combining the atomic gyroscope and the turntable, and it presents a higher sensitivity compared with a combination of the optical gyroscope and the turntable [39]. With the self-calibrated method, the orientation of the gyroscope with respect to the truth north direction was also determined by measuring the absolute Earth's rotation, and its uncertainty is achieved to $0.006$ °. This will provide a prospective supplement to determine the azimuth [40]. Significantly, with the same experimental setup, our results show that the self-calibrated method has better accuracy than the turntable-modulated method. With the scale factor modulated, an arbitrary rotation induced phase shift can be calibrated by determining the period ($2n\pi$) with the VMLF method, which enhances the dynamic range [34]. Therefore, the self-calibrated gyroscope paves a convenient way for measuring an absolute rotation in a wide dynamic range. To achieve a higher accuracy as in the large ring laser gyroscope [7], the uncertainty of the atomic velocity need to be calibrated more precisely. This is possible if considering a large-scale atom interferometer gyroscope ($L = 1.6$ m). In Eq.(4), the distance uncertainty can be further improved to 0.02 ppm if considering the laser interferometric displacement measurement [41], and the phase uncertainty is 0.04 ppm when considering the $1000 : 1$ signal-to-noise level per shot with $T_R=400$ ms and $\delta f=10$ kHz. It is expected that the uncertainty of the Earth rotation measurement reaches $10^{-8}$ order.

In conclusion, we proposed and demonstrated an absolute rotation measurement with a self-calibrated method in a mobile dual-atom-interferometer gyroscope. This self-calibrated atom-interferometer gyroscope is successfully applied to directly measure the Earth's rotation rate to a relative uncertainty of 162 ppm without any auxiliary device (e.g. turntable), and then is verified by determining the true north direction with a precision turntable. This work is the first time to achieve a self-calibration for an atomic gyroscope, which cleverly takes the unique advantage of the atomic speed adjustability. The idea is novel in Sagnac effect based gyroscopes, compared with the similar optical (laser and fiber) gyroscopes where the light speed is unchangeable. Long-term stable and self-calibrated atom-interferometer gyroscopes have important applications in the fields of fundamental physics, geodesy, and long-time inertial navigation.

We acknowledge the financial support from National Innovation Program for Quantum Science and Technology of China under Grant No. 2021ZD0300604, the National Key Research and Development Program of China under Grant No. 2016YFA0302002, the National Natural Science Foundation of China under Grant No. 11674362, No. 91536221, and No. 91736311, the Strategic Priority Research Program of Chinese Academy of Sciences under Grant No. XDB21010100, the Outstanding Youth Foundation of Hubei Province of China under Grant No. 2018CFA082, and the Youth Innovation Promotion Association of Chinese Academy of Sciences.

[1] I. Ciufolini, and E. C. Pavlis, A confirmation of the general relativistic prediction of the Lense-Thirring effect, Nature **431**, 958 (2004).

[2] P. Brosche, and H. Schuh, Tides and earth rotation, Sur. Geophys. **19**, 417 (1998).

[3] H. C. Lefèvre, The fiber-optic gyroscope: Challenges to become the ultimate rotation-sensing technology, Opt. Fib. Techn. **19**, 828 (2013).

[4] A. Gebauer, M. Tercjak, K. U. Schreiber, H. Igel, J. Kodet, U. Hugentobler, J. Wassermann, F. Bernauer, C. J. Lin, S. Donner, S. Egdorf, A. Simonelli, and J. P. R. Wells, Reconstruction of the instantaneous Earth rotation vector with sub-arcsecond resolution using a large scale ring laser array, Phys. Rev. Lett. **125**, 033605 (2020).

[5] C. W. F. Everitt, D. B. DeBra, B. W. Parkinson, J. P. Turneaure, J. W. Conklin et al., Gravity probe B: final results of a space experiment to test general relativity, Phys. Rev. Lett. **106**, 221101 (2011).

[6] F. Bosi, G. Cella, A. D. Virgilio, A. Ortolan, J. P. R. Wells, Measuring gravitomagnetic effects by a multi-ring-laser gyroscope, Phys. Rev. D **84**, 122002 (2011).

[7] R. B. Hurst, M. Mayerbacher, A. Gebauer, K. U. Schreiber, and J.-P. R. Wells, High-accuracy absolute rotation rate measurements with a large ring laser gyro: Establishing the scale factor, Appl. Opt. **56**, 1124 (2017).

[8] A. Tartaglia, A. Di Virgilio, J. Belfi, N. Beverini, and M. L. Ruggiero, Testing general relativity by means of ring lasers, Eur. Phys. J. Plus **132**, 73 (2017).

[9] M. O. Scully, and J. P. Dowling, Quantum-noise limits to matter-wave interferometry, Phys. Rev. A **48**, 3186 (1993).

[10] C. Jentsch, T. Müller, E. M. Rasel, HYPER: A satellite mission in fundamental physics based on high precision atom interferometry, Gen. Rel. Grav. **36**, 2197 (2004).

[11] T. L. Gustavson, A. Landragin, and M.A. Kasevich, Rotation sensing with a dual atom-interferometer Sagnac gyroscope, Class. Quantum Grav. **17**, 2385 (2000).

[12] B. Barrett, R. Geiger, I. Dutta, M. Meunier, B. Canuel, A. Gauguet, P. Bouyer, A. Landragin, The Sagnac effect: 20 years of development in matter-wave interferometry, C. R. Physique **15**, 875 (2014).

[13] M. S. Zhan, J. Wang, W. T. Ni, D. F. Gao, G. Wang, L. X. He, R. B. Li, L. Zhou, X. Chen, J. Q. Zhong, B. Tang, Z. W. Yao, L. Zhu, Z. Y. Xiong, S. B. Lu, G. H. Yu, Q. F. Cheng, M. Liu, Y. R. Liang, P. Xu, X. D. He, M. Ke, Z. Tan, and J. Luo, ZAIGA: Zhaoshan long-baseline atom interferometer gravitation antenna, Int. J. Mod. Phys. D **28**, 1940005 (2020).

[14] R. Gautier, M. Guessoum, L. A. Sidorenkov, Q. Bouton, A. Landragin, R. Geiger, Accurate measurement ofthe Sagnac effect for matter waves, Sci. Adv. **8**, eabn8009 (2022).

[15] B. Canuel, F. Leduc, D. Holleville, A. Gauguet, J. Fils, A. Virdis, A. Clairon, N. Dimarcq, Ch. J. Bordé, A. Landragin and P. Bouyer, Six-axis inertial sensor using cold-atom interferometry, Phys. Rev. Lett. **97**, 010402 (2006).

[16] P. Berg, S. Abend, G. Tackmann, C. Schubert, E. Giese, W. P. Schleich, F. A. Narducci, W. Ertmer, and E. M. Rasel, Composite-light-pulse technique for high-precision atom interferometry, Phys. Rev. Lett. **114**, 063002 (2015).

[17] J. K. Stockton, K. Takase, and M. A. Kasevich, Absolute geodetic rotation measurement using atom interferometry, Phys. Rev. Lett. **107**, 133001 (2011).

[18] Z. W. Yao, S. B. Lu, R. B. Li, J. Luo, J. Wang, and M. S. Zhan, Calibration of atomic trajectories in a large-area dual-atom-interferometer gyroscope, Phys. Rev. A **97**, 013620 (2018).

[19] A. Gauguet, B. Canuel, T. Lefèvre, W. Chaibi, and A. Landragin, Characterization and limits of a cold-atom Sagnac interferometer, Phys. Rev. A **80**, 063604 (2009).

[20] G. Tackmann, P. Berg, C. Schubert, S. Abend, M. Gilowski, W. Ertmer and E. M. Rasel, Self-alignment of a compact large-area atomic Sagnac interferometer, New J. Phys. **14**, 015002 (2012).

[21] Z. W. Yao, H. H. Chen, S. B. Lu, R. B. Li, Z. X. Lu, X. L. Chen, G. H. Yu, M. Jiang, C. Sun, W. T. Ni, J. Wang, and M. S. Zhan, Self-alignment of a large-area dual-atom-interferometer gyroscope using parameter-decoupled phase-seeking calibrations, Phys. Rev. A **103**, 023319 (2021).

[22] I. Dutta, D. Savoie, B. Fang, B. Venon, C. L. G. Alzar, R. Geiger, and A. Landragin, Continuous cold-atom inertial sensor with 1 nrad/sec rotation stability, Phys. Rev. Lett. **116**, 183003 (2016).

[23] D. Savoie, M. Altorio, B. Fang, L. A. Sidorenkov, R. Geiger, and A. Landragin, Interleaved atom interferometry for high-sensitivity inertial measurements, Sci. Adv. **4**, eaau7948 (2018).

[24] M. Altorio, L. A. Sidorenkov, R. Gautier, Accurate trajectory alignment in cold-atom interferometers with separated laser beams, Phys. Rev. A **101**, 033606 (2020).

[25] D. S. Durfee, Y. K. Shaham, and M. A. Kasevich, Long-term stability of an area-reversible atom-interferometer Sagnac gyroscope, Phys. Rev. Lett. **97**, 240801 (2006).

[26] W. J. Xu, L. Cheng, J. Liu, C. Zhang, K. Zhang, Y. Cheng, Z. Gao, L. S. Cao, X. C. Duan, M. K. Zhou, and Z. K. Hu, Effects of wave-front tilt and air density fluctuations in a sensitive atom interferometry gyroscope, Opt. Express. **28**, 12189 (2020).

[27] K. Kotru, D. L. Butts, J. M. Kinast, and R. E. Stoner, Large-area atom interferometry with frequency-swept Raman adiabatic passage, Phys. Rev. Lett. **115**, 103001 (2015).

[28] K. Bongs, M. Holynski, J. Vovrosh, P. Bouyer, G. Condon, E. Rasel, C. Schubert, W. P. Schleich and A. Roura, Taking atom interferometric quantum sensors from the laboratory to real-world applications, Nat. Rev. Phys. **1**, 731 (2019).

[29] C. Jekeli, Navigation error analysis of atom interferometer inertial sensor, Navigation **52**, 1 (2005).

[30] J. Lautier, L. Volodimer, T. Hardin, S. Merlet, M. Lours, F. P. D. Santos, and A. Landragin, Hybridizing matter-wave and classical accelerometers, Appl. Phys. Lett. **105**, 144102 (2014).

[31] L. Zhang, W. Gao, Q. Li, R. B. Li, Z. W. Yao, S. B. Lu, A novel monitoring navigation method for cold atom interference gyroscope, Sensors **19**, 222 (2019).

[32] A. Bonnin, C. Diboune, N. Zahzam, Y. Bidel, M. Cadoret, and A. Bresson, New concepts of inertial measurements with multi-species atom interferometry, Appl. Phys. B **124**, 180 (2018).

[33] C. Avinadav, D. Yankelev, O. Firstenberg, and N. Davidson, Composite-fringe atom interferometry for high-dynamic-range sensing, Phys. Rev. Applied **13**, 054053 (2020).

[34] D. Yankelev, C. Avinadav, N. Davidson, and O. Firstenberg, Atom interferometry with thousand-fold increase in dynamic range, Sci. Adv. **6**, eabd0650 (2020).

[35] C. Antoine, Rotating matter-wave beam splitters and consequences for atom gyrometers, Phys. Rev. A **76**, 033609 (2007).

[36] G. Tackmann, P. Berg, S. Abend, C. Schubert, W. Ertmer, and E. M. Rasel, Large-area Sagnac atom interferometer with robust phase read out, C. R. Physique **15**, 884(2014).

[37] The uncertainty of the scale factor is mainly limited by the atomic velocity because uncertainties of the wave vector ($10^{-9}$) and time ($10^{-8}$) can be ignored in Eq.($2$).

[38] X. W. Zhang, J. Q. Zhong, B. Tang, X. Chen, L. Zhu, P. W. Huang, J. Wang, and M. S. Zhan, Compact portable laser system for mobile cold atom gravimeters, Appl. Opt. **57**, 6545 (2018).

[39] Z. Zhou, Z. W. Tan, X. Y. Wang, and Z. Y. Wang, Experimental analysis of the dynamic north-finding method based on a fiber optic gyroscope, Appl. Opt. **56**, 6504 (2017).

[40] A. Farkas, D. Száz, Á. Egri, M. Blahó, A. Barta, D. Nehéz, B. Bernáth, and G. Horváth, Accuracy of sun localization in the second step of sky-polarimetric Viking navigation for north determination: a planetarium experiment, J. Opt. Soc. Am. A **31**, 1645 (2014).

[41] W. Gao, S. W. Kim, H. Bosse, H. Haitjema, Y. L. Chen, X. D. Lu, W. Knapp, A. Weckenmann, W. T. Estler, and H. Kunzmann, Measurement technologies for precision positioning, CIRP Ann. **64**, 773 (2015).